\documentclass{olplainarticle}
\usepackage{subfigure}
\usepackage{caption}
\usepackage{hyperref}

\title{Curvature Tensor in Discrete Gravity}

\author[1]{Ali H. Chamseddine}
\author[1,2]{Ola Malaeb}
\author[1,2]{Sara Najem}
\affil[1]{Department of Physics, American University of Beirut, Beirut, Lebanon}
\affil[2]{Center for Advanced Mathematical Sciences, American University of Beirut, Beirut, Lebanon}

\begin{abstract}
We study numerically the curvature tensor in a three-dimensional discrete space. Starting from the continuous metric of a three-sphere, we transformed it into a discrete space using three integers $n_1, n_2$, and $n_3$. The numerical results are compared with the expected values in the continuous limit. We show that as the number of cells in the lattice increases, the continuous limit is recovered.  
\end{abstract}

\begin{document}

\flushbottom
\maketitle
\thispagestyle{empty}

\section{Introduction}
The difficulties that arise in attempts to quantize gravity have motivated the development of discrete gravity theories. In discrete gravity, as the naming suggests, the continuous spacetime is discretized and is no longer treated as a smooth manifold, with no unique approach to how the discretization is carried out. For example, lattice gauge theory is the study of gauge theories on a spacetime that has been discretized into a lattice \cite{latticegauge}. Regge calculus is another proposed discrete approximation to general relativity. The formalism involves dividing spacetime into simplices with polyhedrons (Euclidean simplices) as its basic building blocks. Subsequently, the curvature of spacetime is approximated within each simplex \cite{regge}. One other framework is the Euclidean dynamical triangulation (EDT), which approximates the spacetime as a triangulated lattice, where the distances between neighboring points on the lattice are the edges of the triangles \cite{EDT}. One last example is Loop quantum gravity, which is a theory of quantum gravity that postulates that spacetime is represented as a network of finite loops. \cite{LQG}. These approaches are all attempts to develop a theory of gravity that can merge general relativity with quantum mechanics. 

\noindent{} Lately, a new approach of discrete gravity was proposed \cite{discretegravity}. Each elementary cell in the discrete space is completely characterized by displacement operators connecting a cell to its neighbors by the spin connection. The curvature of the discrete space was defined and it was shown that as the elementary volume vanishes, the standard results for the continuous curved differentiable manifolds are fully recovered. 

\noindent{} Recently, and within the lattice-like approach, the scalar curvature of discrete gravity in two dimensions was studied \cite{discrete1}, based on the model proposed in \cite{discretegravity}. Our aim in this paper is to study numerically the curvature tensor in three dimensions in discrete space. In the three-dimensional case, each cell has six neighboring cells which share with it a common boundary. Three integers that can be positive and negative, $n_1, n_2, n_3$, are used to enumerate each of the elementary cells. In the continuous limit, these series of integers become coordinates on the manifold, this is presented in the first section of this paper. In the second section, we convert the continuous metric of a three-sphere into a lattice. We also investigate how the scalar curvature changes in the discrete space depending on the number of cells, and we demonstrate that it approaches the anticipated value in the continuous limit as the number of cells increases.  In the third section, the expected values of the spin connections and the curvature tensor in the continuous case will be compared with the ones acquired in the discrete case.

\section{Isotropic coordinates for a three-sphere}

Starting from a three-sphere metric with a unit radius:
\begin{equation*}
ds^{2} =  d\chi^{2}+\sin^{2}\chi \left(d\theta^2 + sin^2 \theta \ d\phi^{2}\right),
\end{equation*}
we can rewrite it as
\begin{equation*}
ds^{2} =  \frac{dr^{2}}{1-r^2} + r^{2} \left(d\theta^2 + sin^2 \theta \ d\phi^{2}\right),
\end{equation*}
where $r = sin \chi$. 

\begin{figure}[htbp]
\centering
     \includegraphics[scale=1]{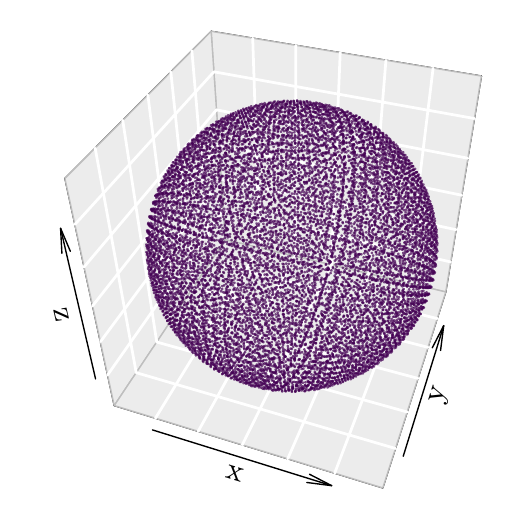}\quad 
   \hfill
   \caption{Three-sphere of radius one formed from the set of the discrete points} \label{3sphereplot}
\end{figure}

\noindent{} To plot the set of discrete points that will make the discrete three-sphere, we start from the regular coordinates given below:
\begin{equation*}
x = r\sin\theta\cos\phi,\quad 
y= r \sin\theta\cos\phi,\quad 
z = r \cos\theta.
\end{equation*}
We define $\left(x_1, x_2, x_3)\right)$ in terms of $n_1$, $n_2$ and $n_3$ and in terms of $\bar{r} = 2 \tan \frac{\chi}{2}$
\begin{equation*}
x_{1} = \frac{2n_{1}}{N} = \bar{r} \sin \theta \cos \phi, \quad x_{2} = \frac{2n_{2}}{N} = \bar{r} \sin \theta \cos \phi, \quad x_{3} = \frac{2n_{3}}{N} = \bar{r} \cos\theta,
\end{equation*}
where 
\begin{equation*}
n_{1}=0,\pm1,\pm2,\cdots,\pm\left(  N-1\right)  ,\quad n_{2}=0,\pm
1,\pm2,\cdots,\pm\left(  N-1\right)  ,\qquad n_{3}=0,\pm1,\pm2,\cdots
,\pm\left(  N-1\right)
\end{equation*}
This finally gives
\begin{equation*}
ds^2 = \frac{d x_1^2 + dx_2^2 + dx_3^2}{\left(1+\frac{\bar{r}^2}{4}\right)^2},
\end{equation*}
with the restriction $n_{1}^{2}+n_{2}^{2}+n_{3}^{2}\leq N^{2}$. The coordinates $(x, y, z)$ will be given by:  
\begin{equation*}
x=\frac{2}{N} \frac{n_{1}}{1+ \frac{\left(  n_{1}^{2}+n_{2}^{2}+n_3^2\right)} {N^{2}}},\quad
y=\frac{2}{N} \frac{n_{2}}{1+ \frac{\left(  n_{1}^{2}+n_{2}^{2}+n_3^2\right)} {N^{2}}},\quad
z=\frac{2}{N} \frac{n_{3}}{1+ \frac{\left(  n_{1}^{2}+n_{2}^{2}+n_3^2\right)} {N^{2}}}
\end{equation*}
The set of discrete points forming a three-sphere of radius one are displayed in Figure \ref{3sphereplot}.

\section{Three Dimensional Lattice Gravity} 

This case is simple because the rotation group $SO\left(  3\right)$ has the same Lie Algebra as $SU\left(  2\right)$. We take the connection:
\begin{equation*}
\frac{1}{4}\omega_{\mu}^{ij}\left(  n\right)  \gamma_{ij}\equiv\frac{i}%
{2}\omega_{\mu}^{i}\left(  n\right)  \sigma_{i}\quad\omega_{\mu}^{i}\left(
n\right)  =\frac{1}{2}\epsilon^{ijk}\omega_{\mu}^{\quad jk}\left(  n\right).
\end{equation*}
Thus the curvature $R_{\mu\nu}^{\quad ij}\left(  n\right)  =\epsilon
^{ijk}R_{\mu\nu}^{\quad k}\left(  n\right)  $ is given by (no summation on $\mu,\nu$ in what follows)
\begin{equation}
R_{\mu\nu}^{\quad i}\left(  n\right)  =\frac{2}{\mathcal{\ell}^{\mu
}\mathcal{\ell}^{\nu}}\left(  A_{\mu\nu}\left(  n\right)  B_{\mu\nu}%
^{i}\left(  n\right)  -A_{\nu\mu}\left(  n\right)  B_{\nu\mu}^{i}\left(
n\right)  +\epsilon^{ijk}B_{\mu\nu}^{j}\left(  n\right)  B_{\nu\mu}^{k}\left(
n\right)  \right), 
\end{equation} \label{curvature}
where
\begin{equation}
A_{\mu\nu}\left(  n\right)  =\left(  \cos\frac{1}{2}\mathcal{\ell}\omega_{\mu
}\left(  n+\widehat{\nu}\right)  \cos\frac{1}{2}\mathcal{\ell}\omega_{\nu
}\left(  n\right)  -\widehat{\omega}_{\mu}^{j}\left(  n+\widehat{\nu}\right)
\widehat{\omega}_{\nu}^{j}\left(  n\right)  \sin\frac{1}{2}\mathcal{\ell
}\omega_{\mu}\left(  n+\widehat{\nu}\right)  \sin\frac{1}{2}\mathcal{\ell
}\omega_{\nu}\left(  n\right)  \right),
\end{equation} \label{Amunu}
and
\begin{align}
B_{\mu\nu}^{i}\left(  n\right)   &  =\left(  \widehat{\omega}_{\mu}^{i}\left(
n\right)  \sin\frac{1}{2}\mathcal{\ell}\omega_{\mu}\left(  n\right)  \cos
\frac{1}{2}\mathcal{\ell}\omega_{\nu}\left(  n+\widehat{\mu}\right)
+\widehat{\omega}_{\nu}^{i}\left(  n+\widehat{\mu}\right)  \sin\frac{1}%
{2}\mathcal{\ell}\omega_{\nu}\left(  n+\widehat{\mu}\right)  \cos\frac{1}%
{2}\mathcal{\ell}\omega_{\mu}\left(  n\right)  \right. \nonumber\\
&  \left.  -\epsilon^{ijk}\widehat{\omega}_{\mu}^{j}\left(  n\right)
\sin\frac{1}{2}\mathcal{\ell}\omega_{\mu}\left(  n\right)  \widehat{\omega
}_{\nu}^{k}\left(  n+\widehat{\mu}\right)  \sin\frac{1}{2}\mathcal{\ell}%
\omega_{\nu}\left(  n+\widehat{\mu}\right)  \right).
\end{align} \label{Bmunu}
We have denoted
\begin{equation}
\quad\left(  \omega_{\mu}\left(  n\right)  \right)  ^{2}\equiv%
{\displaystyle\sum\limits_{i=1}^{3}}
\omega_{\mu}^{i}\left(  n\right)  \omega_{\mu}^{i}\left(  n\right)
,\quad\widehat{\omega}_{\mu}^{i}\left(  n\right)  \equiv\frac{\omega_{\mu}%
^{i}\left(  n\right)  }{\omega_{\mu}\left(  n\right)  }
\end{equation} \label{definition}
The torsion is given by
\begin{align*}
T_{\mu\nu}^{\quad i}\left(  n\right)   &  =\frac{1}{\mathcal{\ell}^{\mu}%
}\left(  \cos\mathcal{\ell}\omega_{\mu}\left(  n\right)  e_{\nu}^{i}\left(
n+\widehat{\mu}\right)  -\epsilon^{ijk}\sin\mathcal{\ell}\omega_{\mu} \left(n\right)  \widehat{\omega}_{\mu}^{j}\left(  n\right)  e_{\nu}^{k}\left(
n+\widehat{\mu}\right)  \right. \nonumber\\
&  \left.  +2\widehat{\omega}_{\mu}^{i}\left(  n\right)  \widehat{\omega}%
_{\mu}^{j}\left(  n\right)  \sin^{2}\frac{1}{2}\mathcal{\ell}\omega_{\mu
}\left(  n\right)  e_{\nu}^{j}\left(  n+\widehat{\mu}\right)  -e_{\nu}%
^{i}\left(  n\right)  \right)  -\left(  \mu\leftrightarrow\nu\right)
\end{align*}
The vanishing of $T_{\mu\nu}^{\quad i}$ provides nine conditions to solve for the nine unknowns $\omega_{\mu}^{i}\left(  n\right)$ (check appendix I). We then proceed to find the components of the curvature $R_{\mu \nu}^{i}$ using equations \ref{curvature}, \ref{Amunu} and \ref{Bmunu} (check appendix II). 

\noindent{} The curvature tensor is given by
\begin{equation*}
R_{\mu\nu}^{\quad ij}=\epsilon^{ijk}R_{\mu\nu}^{k},%
\end{equation*}
and the scalar curvature is
\begin{equation*}
R=%
{\displaystyle\sum\limits_{\mu,\nu}}
e_{i}^{\mu}e_{j}^{\nu}R_{\mu\nu}^{\quad ij}=%
{\displaystyle\sum\limits_{\mu,\nu}}
\epsilon^{ijk}e_{i}^{\mu}e_{j}^{\nu}R_{\mu\nu}^{k}%
\end{equation*}
To show that the curvature tensor has the correct continuous limit, we note that as $\mathcal{\ell}\rightarrow0$
\begin{align*}
A_{\mu\nu}\left(  n\right)   &  \rightarrow1+O\left(  \mathcal{\ell}%
^{2}\right) \\
B_{\mu\nu}^{i}\left(  n\right)   &  \rightarrow\frac{\mathcal{\ell}}{2}\left(
\omega_{\mu}^{i}\left(  n\right)  +\omega_{\nu}^{i}\left(  n+\widehat{\mu
}\right)  \right)  -\frac{\mathcal{\ell}^{2}}{4}\left(  \epsilon^{ijk}%
\omega_{\mu}^{j}\left(  n\right)  \omega_{\nu}^{k}\left(  n\right)  \right)
+O\left(  \mathcal{\ell}^{3}\right)
\end{align*}
This implies that
\begin{align*}
R_{\mu\nu}^{\quad i}\left(  n\right)   &  \rightarrow\frac{2}{\mathcal{\ell
}^{2}}\frac{\mathcal{\ell}}{2}\left(  \omega_{\mu}^{i}\left(  n\right)
+\omega_{\nu}^{i}\left(  n+\widehat{\mu}\right)  -\omega_{\nu}^{i}\left(
n\right)  -\omega_{\mu}^{i}\left(  n+\widehat{\nu}\right)  \right)  +\left(
\frac{2}{\mathcal{\ell}^{2}}\right)  \left(  -\frac{\mathcal{\ell}^{2}}%
{4}\right)  2\epsilon^{ijk}\omega_{\mu}^{j}\left(  n\right)  \omega_{\nu}%
^{k}\left(  n\right) \nonumber\\
&  \rightarrow\partial_{\mu}\omega_{\nu}^{i}-\partial_{\nu}\omega_{\mu}%
^{i}-\epsilon^{ijk}\omega_{\mu}^{j}\omega_{\nu}^{k},%
\end{align*}
where we have used $\omega_{\nu}^{i}\left(  n+\widehat{\mu}\right)
-\omega_{\nu}^{i}\left(  n\right)  =\mathcal{\ell\Delta}_{\mu}\omega_{\nu}%
^{i}+O\left(  \mathcal{\ell}^{2}\right)  .$ Noting that
\begin{equation*}
R_{\mu\nu}^{\quad ij}=\epsilon^{ijk}R_{\mu\nu}^{k}=\partial_{\mu}\omega_{\nu}^{ij}-\partial_{\nu}\omega_{\mu}^{ij} + \omega_{\mu}^{ik}\omega_{\nu}^{kj}-\omega_{\nu}^{ik}\omega_{\mu}^{kj},
\end{equation*}
where $\omega_{\mu}^{ij} = \epsilon^{ijk}\omega_{\mu}^{k}$ recovers the familiar expression for the curvature tensor.

\noindent{} In the case when $e_{\mu}^{i}$ is diagonal then
\begin{equation*}
R=2\left(  \frac{1}{e_{1}e_{2}}R_{\overset{.}{1}\overset{.}{2}}^{3}+\frac
{1}{e_{2}e_{3}}R_{\overset{.}{2}\overset{.}{3}}^{1}+\frac{1}{e_{3}e_{1}%
}R_{\overset{.}{3}\overset{.}{1}}^{2}\right)
\end{equation*}
Considering the example of the three-sphere in isotropic coordinates we will have
\begin{equation*}
e_{\overset{.}{1}}^{1}\left(  n\right)  =e_{\overset{.}{2}}^{2}\left(
n\right)  =e_{\overset{.}{3}}^{3}\left(  n\right)  =e\left(  n\right)
=\frac{1}{1+\frac{1}{N^{2}}\left(  n_{1}^{2}+n_{2}^{2}+n_{3}^{2}\right)  }%
\end{equation*}
Since $\left(  x^{1}\right)  ^{2}+\left(  x^{2}\right)  ^{2}+\left(
x^{3}\right)  ^{2}=r^{2}=2\tan\frac{\theta}{2}$, one must use two coordinate charts. Therefore, one must have $r^{2}\leq4$ and thus $\left(  n^{1}\right)
^{2}+\left(  n^{2}\right)  ^{2}+\left(  n^{3}\right)  ^{2}\leq N^{2}.$ This restriction is imposed on the values of $n_1, n_2, n_3$.

\section{Continuous case}

It is helpful to list the values of the spin connections and curvatures in the
continuous limit. In a numerical calculation to test the accuracy of the program, we can compare the
numerical values of $\omega_{\mu}^{i}$, $R_{\mu\nu}^{\quad ij}$ to those of the continuous limit. In particular, one can test the homogeneity of the space by comparing the Ricci tensor $R_{\nu\sigma}$ to the metric $g_{\nu\sigma}$
\begin{equation*}
R_{\nu\sigma}=e_{i}^{\mu}R_{\mu\nu}^{\quad ij}e_{\sigma j}=2g_{\nu\sigma}%
\end{equation*}

\begin{equation*}
R_{\mu\nu}^{\quad ij}=e_{\mu}^{k}e_{\nu}^{l}R_{kl}^{\quad ij}=\left(  e_{\mu
}^{i}e_{\nu}^{j}-e_{\mu}^{j}e_{\nu}^{i}\right)  =\epsilon^{ijk}R_{\mu\nu}^{k}%
\end{equation*}
Inverting we get
\begin{equation*}
R_{\mu\nu}^{k}=\epsilon^{kij}e_{\mu}^{i}e_{\nu}^{j},
\end{equation*}
and the only non-vanishing curvatures are:
\begin{equation*}
R_{\overset{.}{1}\overset{.}{2}}^{\quad12}=R_{\overset{.}{1}\overset{.}{2}%
}^{3}=\frac{1}{\left(  1+\frac{1}{4}r^{2}\right)  ^{2}}=R_{\overset{.}{2}%
\overset{.}{3}}^{1}=R_{\overset{.}{3}\overset{.}{1}}^{2}%
\end{equation*}
The total curvature scalar is then
\begin{equation*}
R=2\left(  R_{\overset{.}{1}\overset{.}{2}}^{\quad\overset{.}{1}%
\overset{.}{2}}+R_{\overset{.}{2}\overset{.}{3}}^{\quad\overset{.}{2}%
\overset{.}{3}}+R_{\overset{.}{1}\overset{.}{3}}^{\quad\overset{.}{1}%
\overset{.}{3}}\right)  =6
\end{equation*}
We also note that the continuous limit of $\omega_{\mu}^{i}=\frac{1}%
{2}\epsilon^{ijk}\omega_{\mu}^{jk}=\frac{1}{2}\epsilon^{ijk}x^{j}e_{\mu}^{k}.$
In components we have
\begin{align*}
\omega_{\overset{.}{1}}^{1} &  =0,\quad\omega_{\overset{.}{2}}^{1}=-\frac
{1}{2}x^{3}e,\quad\omega_{\overset{.}{3}}^{1}=\frac{1}{2}x^{2}e\\
\omega_{\overset{.}{1}}^{2} &  =\frac{1}{2}x^{3}e,\quad\omega_{\overset{.}{2}%
}^{2}=0,\quad\omega_{\overset{.}{3}}^{2}=-\frac{1}{2}x^{1}e\\
\omega_{\overset{.}{1}}^{3} &  =-\frac{1}{2}x^{2}e,\quad\omega_{\overset{.}{2}%
}^{3}=\frac{1}{2}x^{1}e,\quad\omega_{\overset{.}{3}}^{3}=0
\end{align*}
The expression appearing in the numerical $\mathcal{\ell\omega}_{\mu}$ must
correspond to
\begin{align*}
\left(  \omega\overset{.}{_{1}}\right)  ^{2} &  =\frac{1}{4}\left(  \left(
x^{2}\right)  ^{2}+\left(  x^{3}\right)  ^{2}\right)  e^{2},\\
\left(  \omega\overset{.}{_{2}}\right)  ^{2} &  =\frac{1}{4}\left(  \left(
x^{1}\right)  ^{2}+\left(  x^{3}\right)  ^{2}\right)  e^{2},\\
\left(  \omega\overset{.}{_{3}}\right)  ^{2} &  =\frac{1}{4}\left(  \left(
x^{1}\right)  ^{2}+\left(  x^{2}\right)  ^{2}\right)  e^{2},
\end{align*}
so that
\begin{align*}
\widehat{\omega}_{\overset{.}{1}}^{1} &  =0, \quad\widehat{\omega}_{\overset{.}{1}}^{2} = \frac{x^{3}}{\sqrt{\left(  x^{2}\right)  ^{2} + \left(x^{3}\right)^{2}}}, \quad\widehat{\omega}_{\overset{.}{1}}^{3} = -\frac{x^{2}}{\sqrt{\left( x^{2}\right)^{2} + \left( x^{3}\right)  ^{2}}} \\
\widehat{\omega}_{\overset{.}{2}}^{1} &  =-\frac{x^{3}}{\sqrt{\left(x^{1}\right) ^{2} + \left(  x^{3}\right)^{2}}}, \quad\widehat{\omega
}_{\overset{.}{2}}^{2}=0, \quad\widehat{\omega}_{\overset{.}{2}}^{3}
=\frac{x^{1}}{\sqrt{\left( x^{1}\right)  ^{2} +\left(  x^{3}\right)  ^{2}}}\\
\widehat{\omega}_{\overset{.}{3}}^{1} &  =\frac{x^{2}}{\sqrt{\left(x^{1}\right)  ^{2} +\left(  x^{2}\right)^{2}}}, 
 \quad\widehat{\omega}_{\overset{.}{3}}^{2} = -\frac{x^{1}}{\sqrt{\left(  x^{1}\right)  ^{2}+\left(x^{2}\right)^{2}}}, \quad\widehat{\omega}_{\overset{.}{3}}^{3}=0.
\end{align*}
To go to discretized form we let $x^{i}=2\frac{n^{i}}{N}$ so that
${\frac{\mathcal{\ell}}{2} \omega}_{i}^{j}\rightarrow\frac{\epsilon^{ijk}n^{k}}%
{1+\frac{n^{l}n^{l}}{N^{2}}}$ provided we take $\mathcal{\ell=}\frac{2}{N}.$ 

\begin{figure}[htbp]
\centering
     \includegraphics[scale=1]{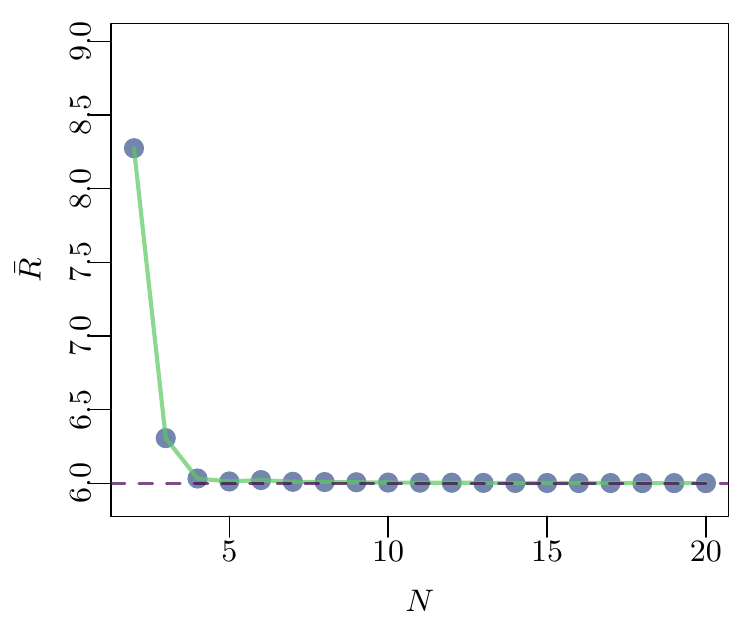}\quad  
   \caption{The mean value of the curvature $\bar{R}$ for different values of $N$.} \label{meanR}
\end{figure}

\begin{figure}[htbp]
     \includegraphics[scale=0.72]{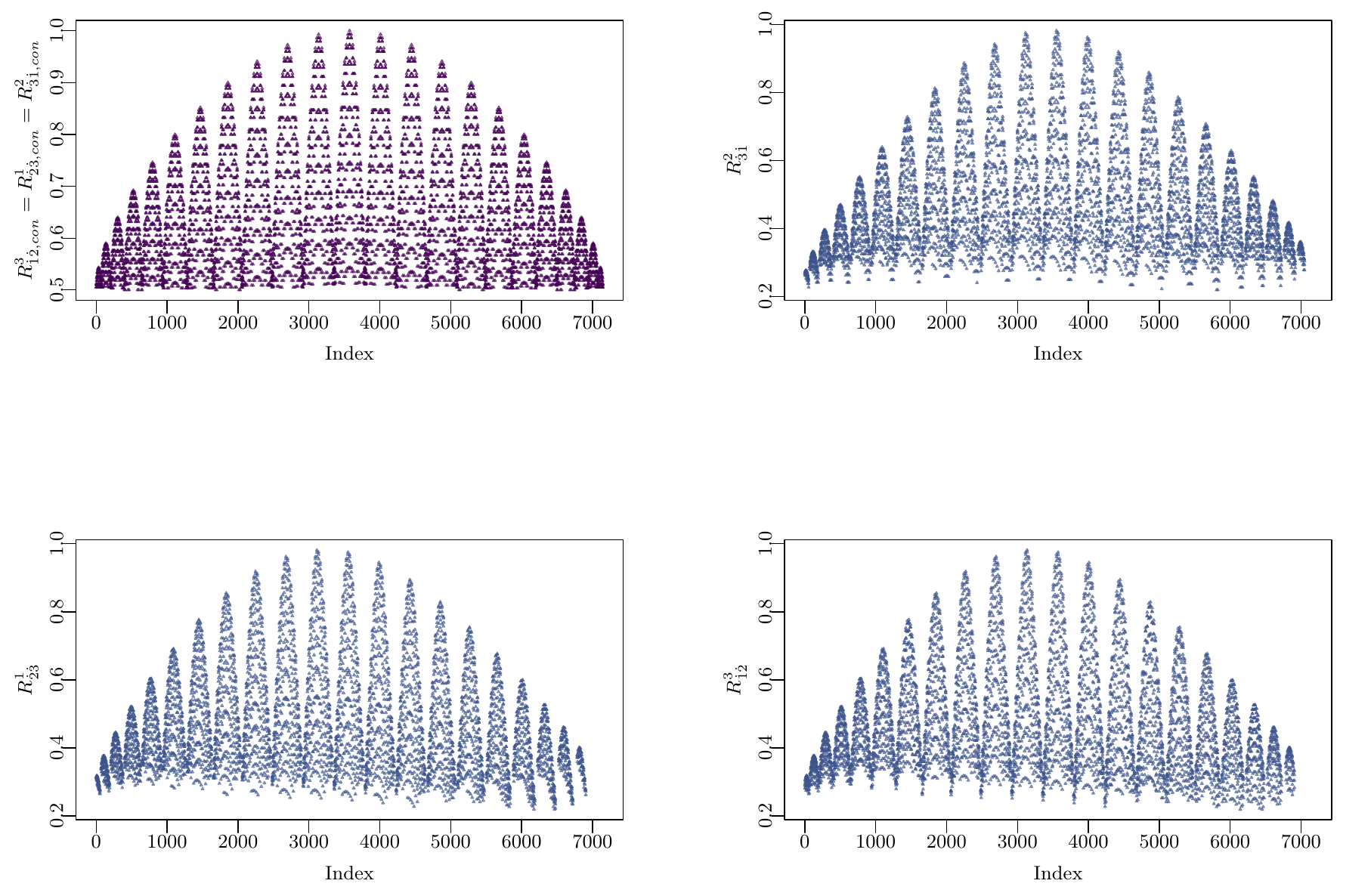}\quad  
   \caption{$R_{\dot{1} \dot{2}, con}^{3}$, $R_{\dot{2} \dot{3},con}^{1}$ and $R_{\dot{3} \dot{1},con}^{2}$ in the continuous case (which are equal) are compared to $R_{\dot{1} \dot{2}}^{3}$, $R_{\dot{2} \dot{3}}^{1}$, and $R_{\dot{3} \dot{1}}^{2}$ in the discrete case} \label{RS}
\end{figure}

\begin{figure}[htbp]
     \includegraphics[scale=0.72]{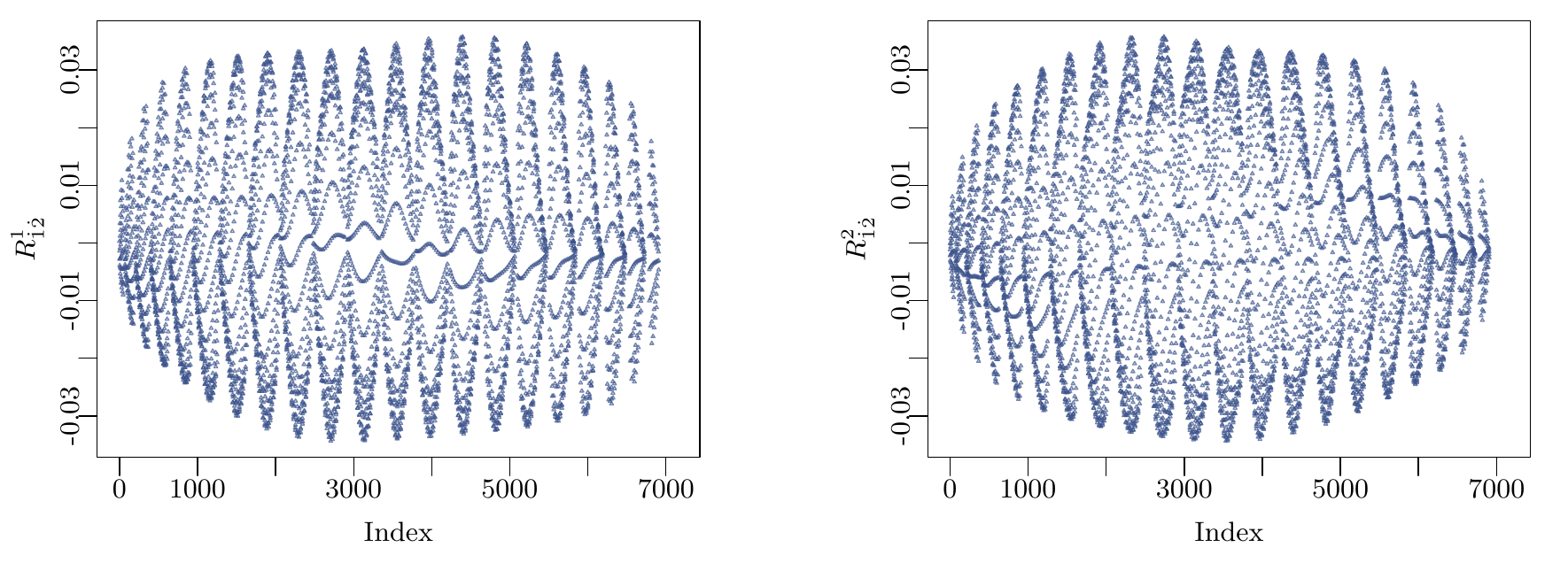}\quad  \caption{The values of $R_{\dot{1} \dot{2}}^{1}$ and $R_{\dot{1} \dot{2}}^{2}$ in the discrete case are close to zero as expected} \label{R112R211}
\end{figure}

\begin{figure}[htbp]
     \includegraphics[scale=0.72]{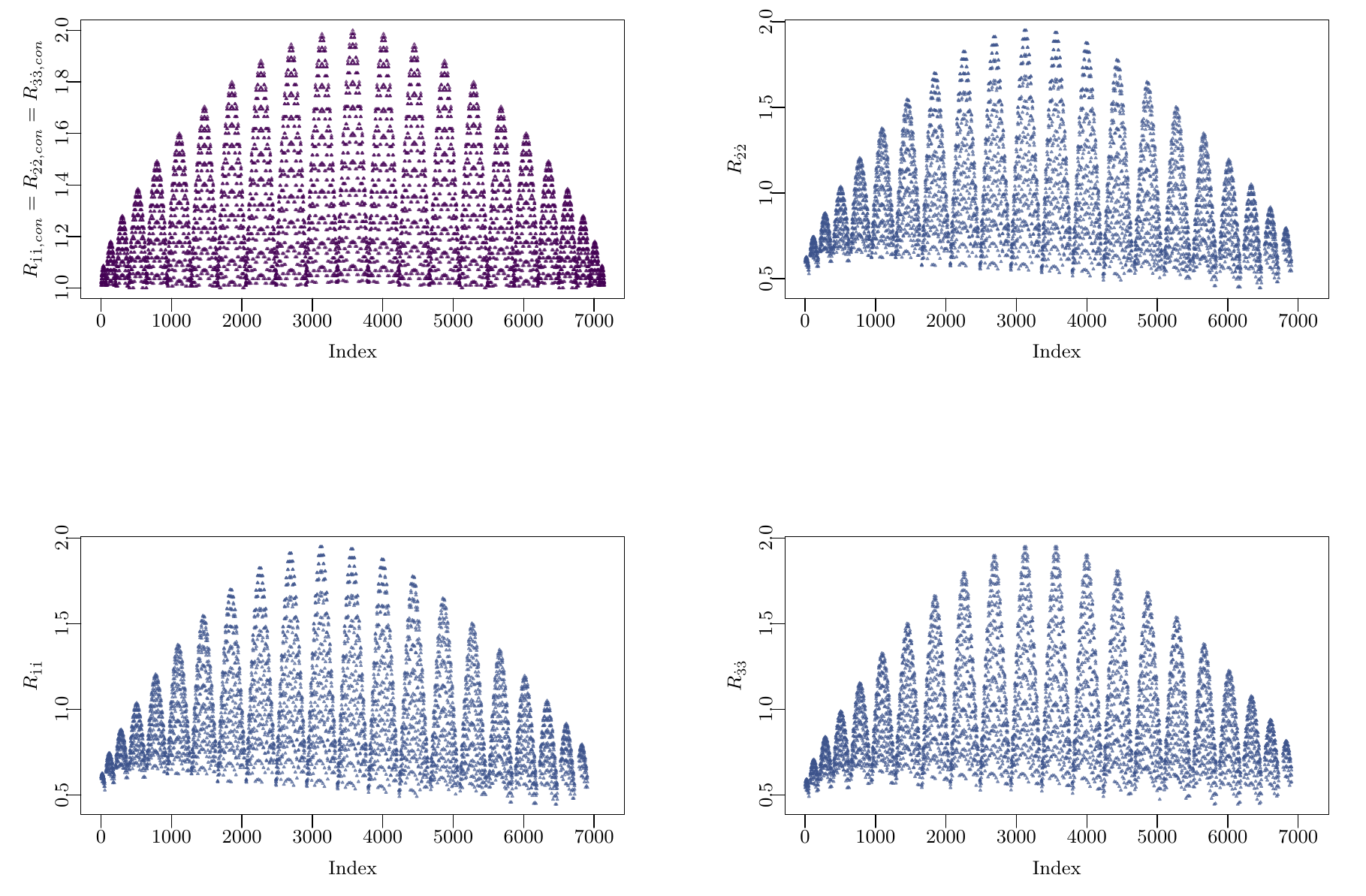}\quad  
   \caption{$R_{\dot{1} \dot{1}, con}$ = $R_{\dot{2} \dot{2}, con}$ = $R_{\dot{3} \dot{3}, con}$ in the continuous case are compared to $R_{\dot{1} \dot{1}}$, $R_{\dot{2} \dot{2}}$, and $R_{\dot{3} \dot{3}}$ in the discrete case} \label{R112233}
\end{figure}

\begin{figure}[htbp]
     \includegraphics[scale=0.72]{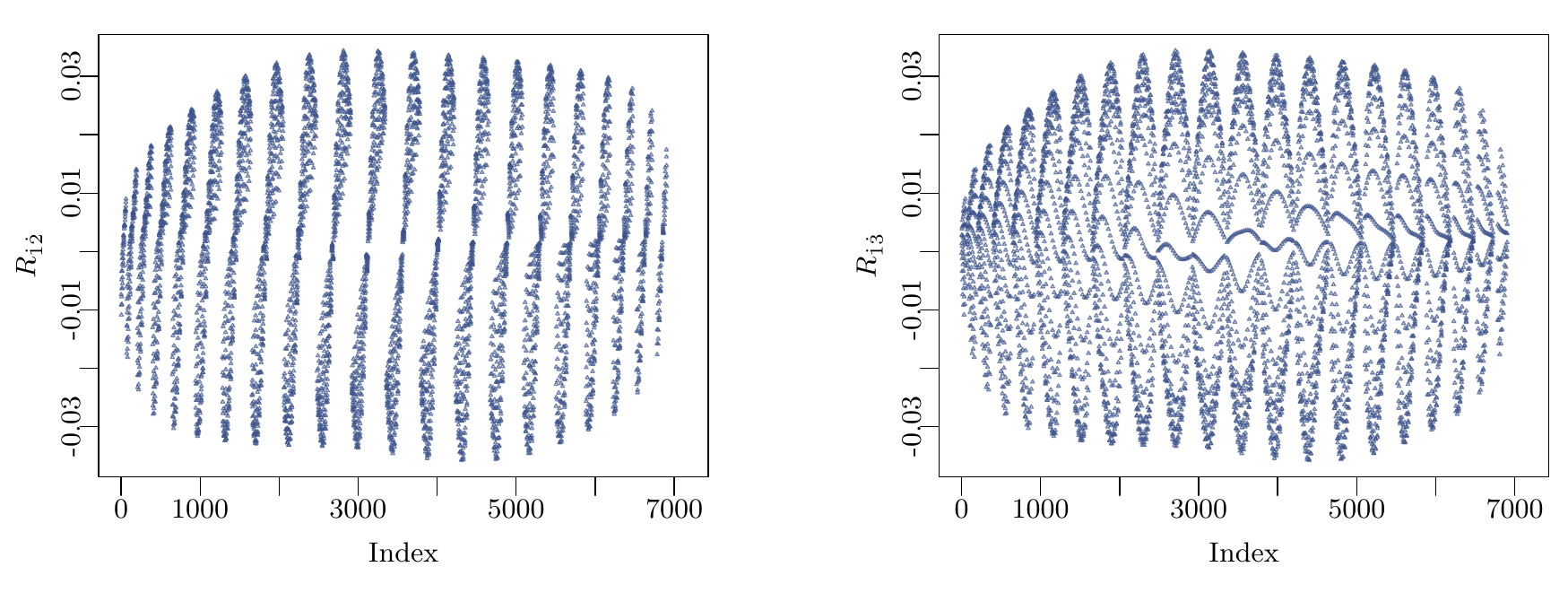}\quad  
   \caption{The values of $R_{\dot{1} \dot{2}}$ and $R_{\dot{1} \dot{3}}$ in the discrete case are close to zero as expected} \label{R12R13}
\end{figure}

\begin{figure}[htbp]
     \includegraphics[scale=1]{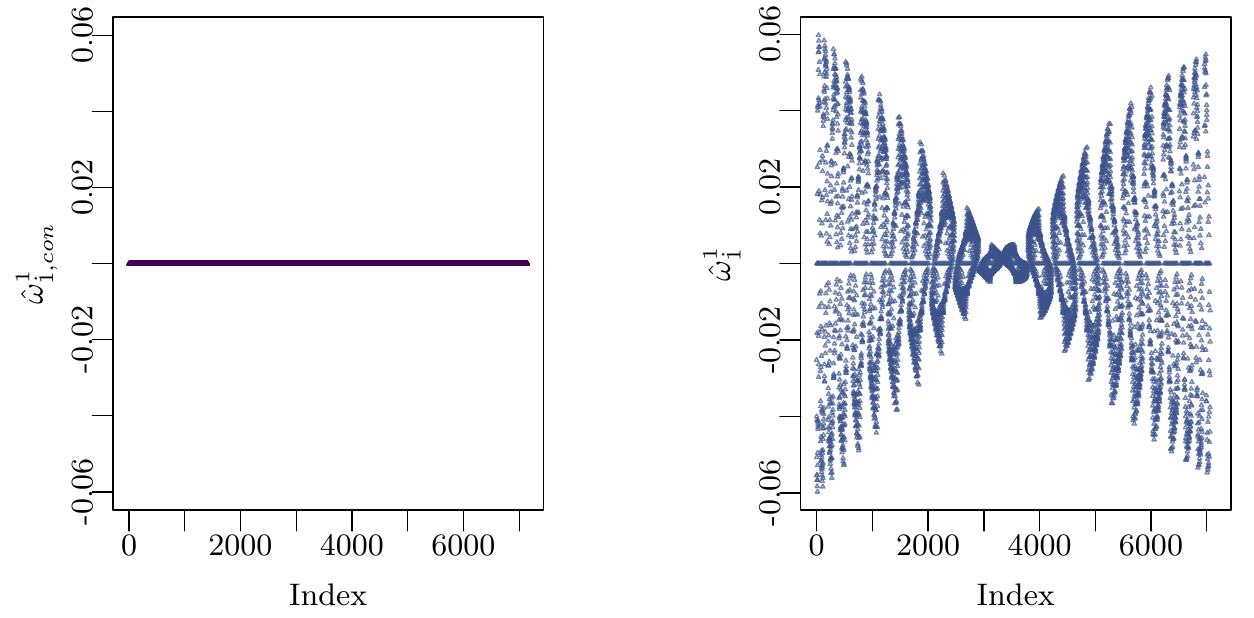}\quad  
   \caption{${\widehat{\omega}}^1_{\dot{1},con}$ in the continuous limit is compared to ${\widehat{\omega}}^1_{\dot{1}}$ in the discrete case} \label{n12x1}
\end{figure}

\begin{figure}[htbp]
     \includegraphics[scale=1]{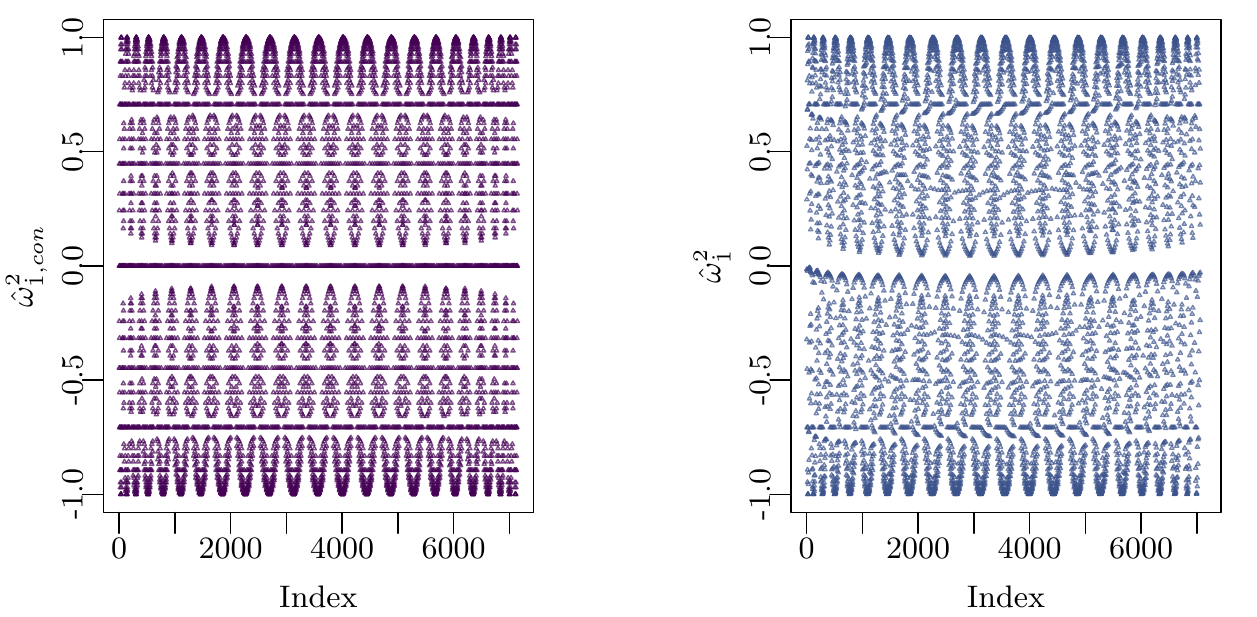}\quad  
 
 \caption{${\widehat{\omega}}^2_{\dot{1},con}$ in the continuous limit is compared to ${\widehat{\omega}}^2_{\dot{1}}$ in the discrete case} \label{n12x4}
\end{figure}

\begin{figure}[htbp]
     \includegraphics[scale=1]{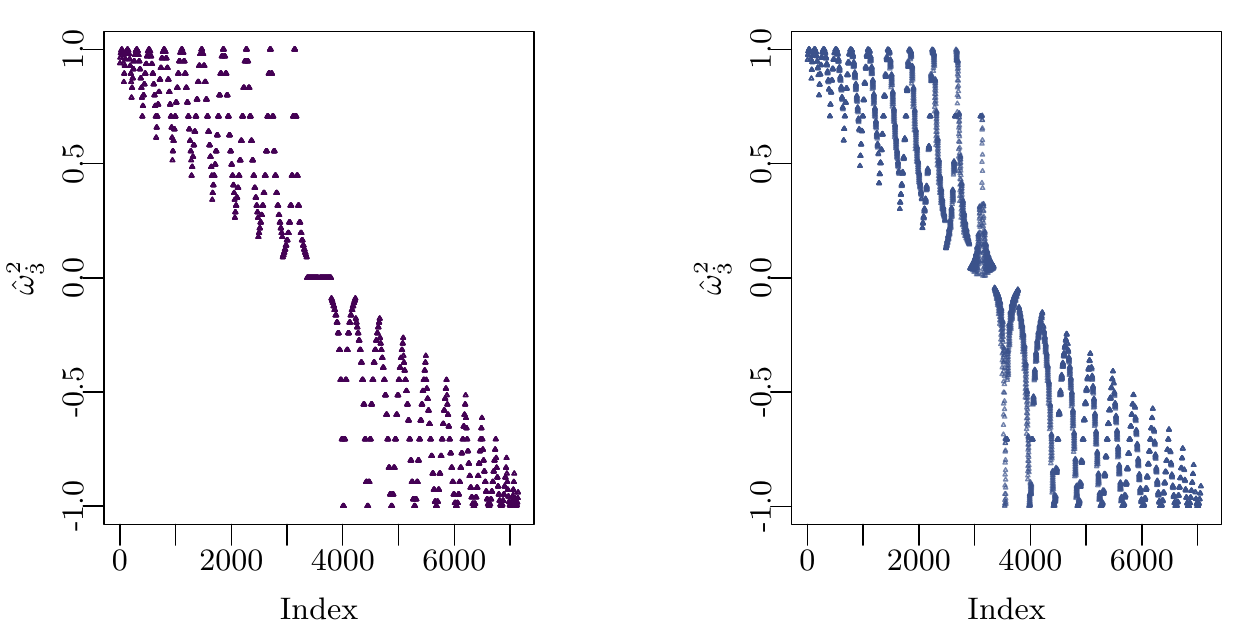}\quad  
   \caption{${\widehat{\omega}}^2_{\dot{3},con}$ in the continuous limit is compared to ${\widehat{\omega}}^2_{\dot{3}}$ in the discrete case} \label{n12x6}
\end{figure}

\begin{figure}[htbp]
     \includegraphics[scale=1]{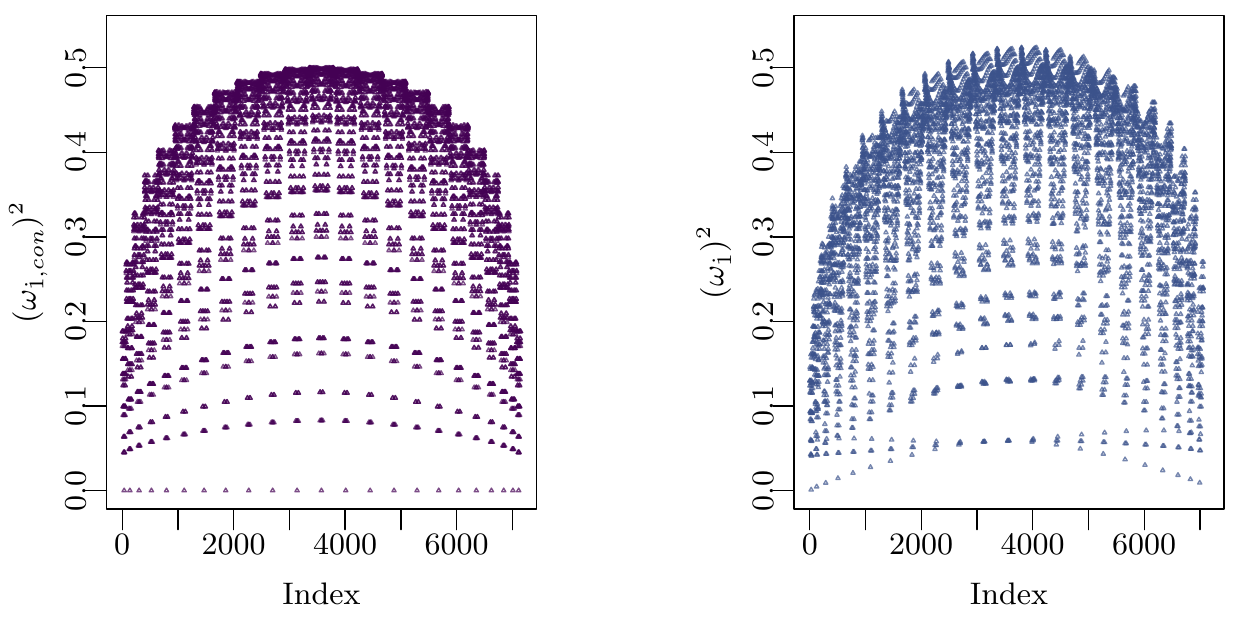}\quad  
   \caption{${\omega}_{\dot{1}}$ in the continuous limit compared to ${\omega}_{\dot{1}}$ in the discrete case} \label{n12y1}
\end{figure}

\begin{figure}[htbp]
     \includegraphics[scale=1]{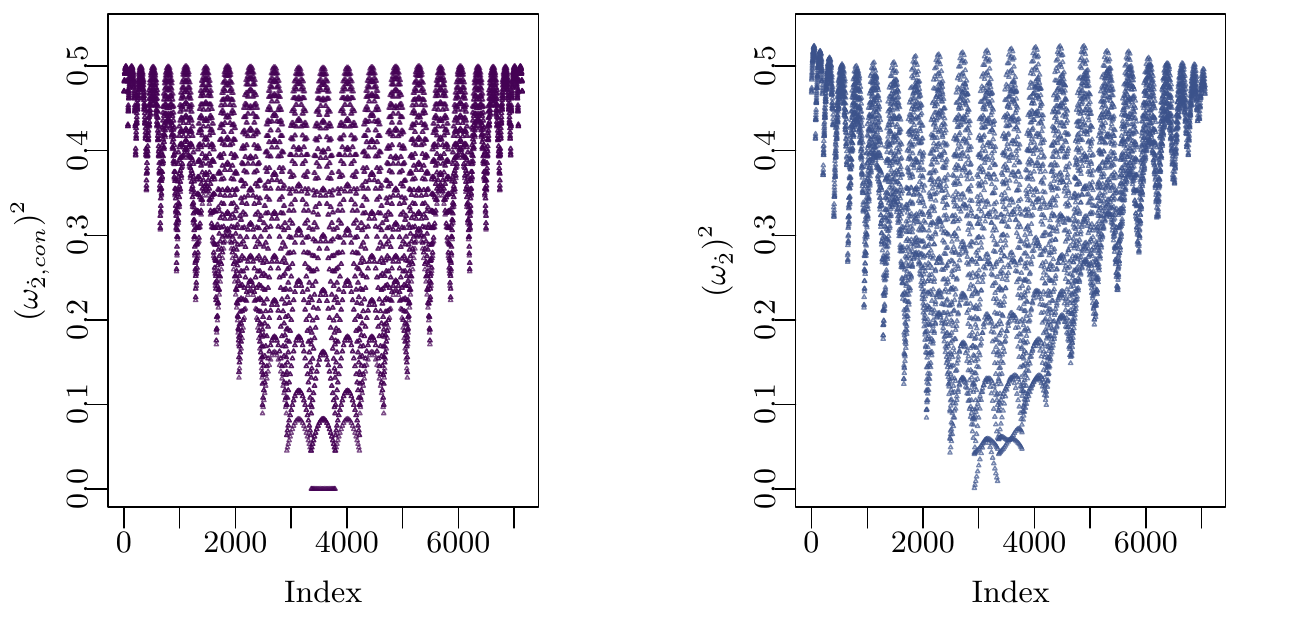}\quad  
   \caption{${\omega}_{\dot{2},con}$ in the continuous limit is compared to ${\omega}_{\dot{2}}$ in the discrete case} \label{n12y2}
\end{figure}

\begin{figure}[htbp]
     \includegraphics[scale=1]{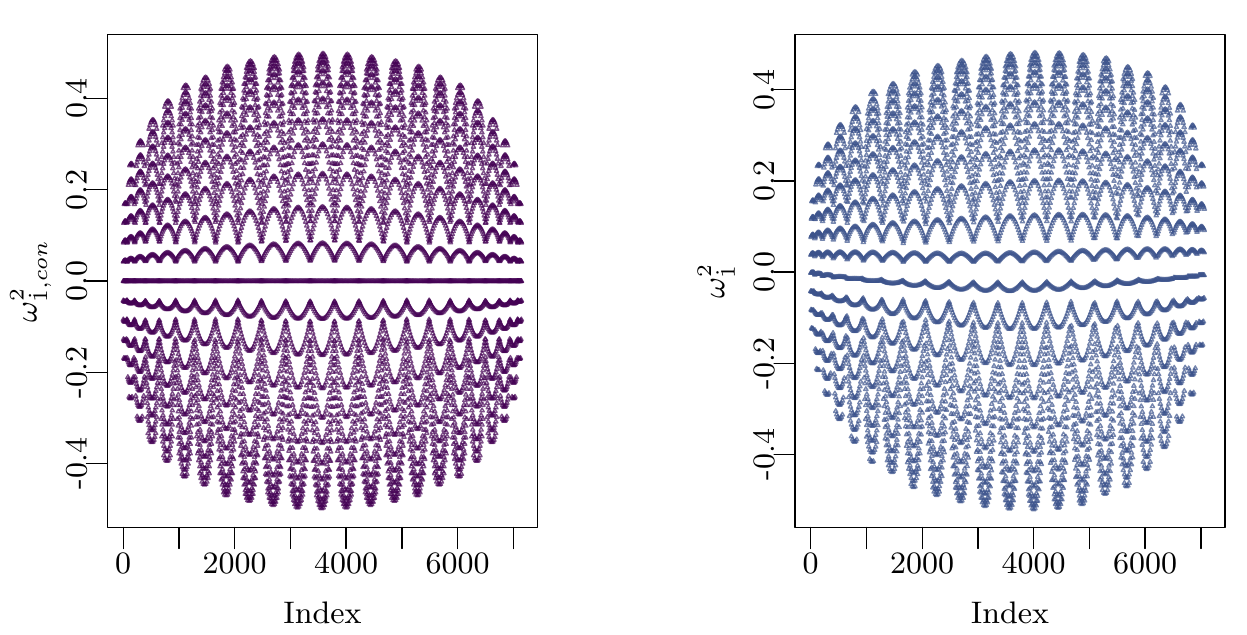}\quad  
   \caption{${{\omega}}^2_{\dot{1},con}$ in the continuous limit is compared to ${{\omega}}^2_{\dot{1}}$ in the discrete case} \label{n12z2}
\end{figure}

\begin{figure}[htbp]
     \includegraphics[scale=1]{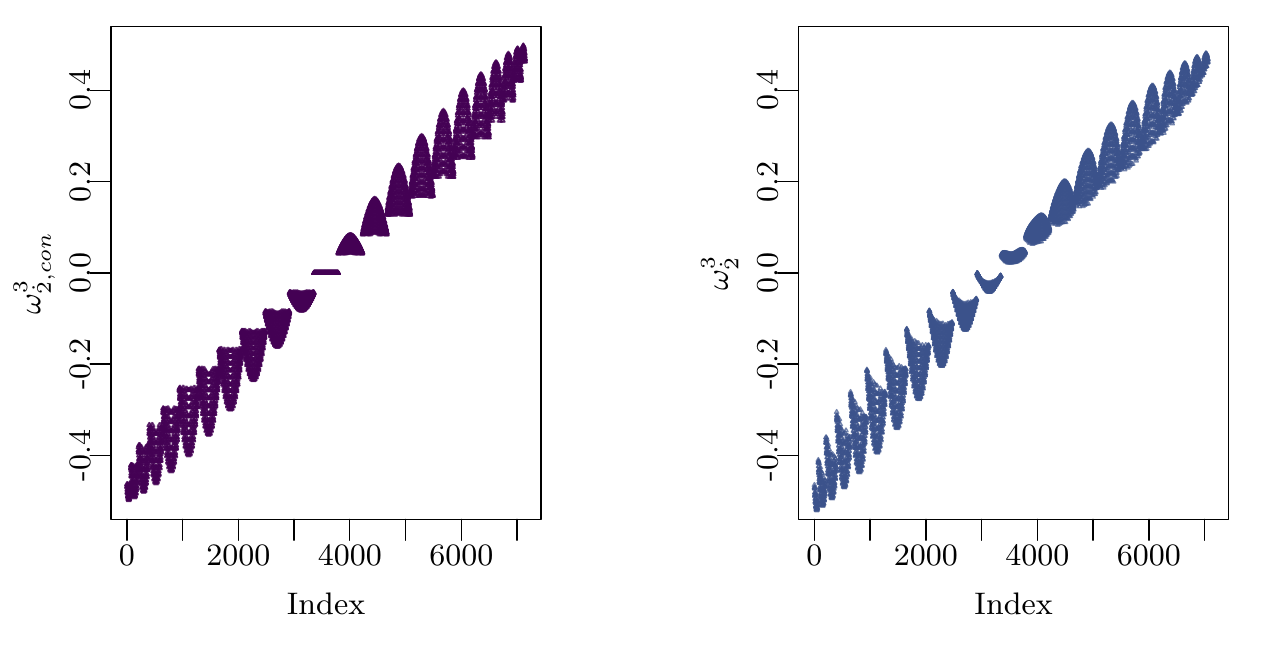}\quad  
   \caption{${{\omega}}^3_{\dot{2},con}$ in the continuous limit is compared to ${{\omega}}^3_{\dot{2}}$ in the discrete case} \label{n12z6}
\end{figure}

Figure \ref{meanR} illustrates the convergence of the mean curvature as a function of $N$ to its expected value of $6$. Figures \ref{RS} and \ref{R112R211} show the curvature in the continuous and discrete limit. In the continuous limit, $R_{\dot{1} \dot{2}}^{3}$ has the same value as $R_{\dot{2} \dot{3}}^{1}$ and $R_{\dot{3} \dot{1}}^{2}$. Their plot is compared with each of the other plots in the discrete case in Figure \ref{RS}. While Figure \ref{R112R211} presents $R_{\dot{1} \dot{2}}^{1}$ and $R_{\dot{1} \dot{2}}^{2}$ that vanish in the continuous case. Also, the Ricci curvature is plotted in Figures \ref{R112233} and \ref{R12R13}. In Figure \ref{R112233}, the values of $R_{\dot{1}\dot{1}}$, $R_{\dot{2}\dot{2}}$, and $R_{\dot{2}\dot{2}}$ are equal in the continuous case and are compared to their values in the discrete case. While Figure \ref{R12R13} shows the plots of $R_{\dot{1}\dot{2}}$ and $R_{\dot{1}\dot{3}}$ in the discrete case. The latter is expected to vanish in the continuous case. To compare the numerical values of the spin connection, Figures \ref{n12x1}-\ref{n12z6} are represented. These figures are a sample of the spin connections plotted in the discrete case versus the plots of the expected values in the continuous case, for $N=12$. The plots in the discrete case were obtained by solving Equations \ref{one}-\ref{nine} combined with Equation \ref{definition} using \textit{nleqslv}, which is an \textit{R} package for solving nonlinear equations.  

\clearpage

\begin{figure}[htbp]
     \includegraphics[scale=0.7]{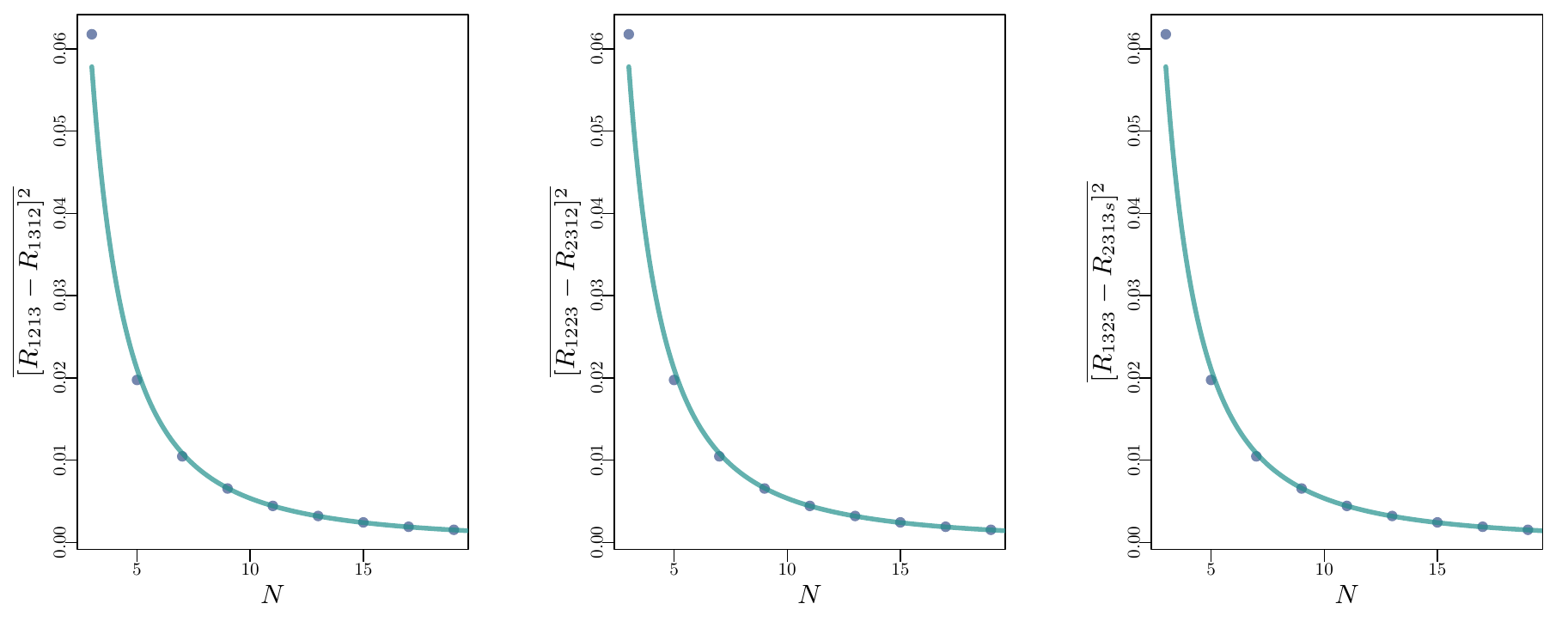}\quad  
   \caption{The Root-mean-squared error of $R_{1213} - R_{1312}$, $R_{1223} - R_{2312}$ and $R_{1323} - R_{2313}$ versus $N$ is shown.} \label{R4indices}
\end{figure}

\noindent Having established the rapid convergence of the components of the Riemann tensor to the continuous limit, we examine the properties of the discrete values of the tensors and whether they exhibit the same symmetries as the continuous case. We thus define:
\begin{equation*}
    R_{kl}^{\quad ij} = \sum_{\mu = 1}^3 \sum_{\nu = 1}^3 e_k^{\mu}e_l^{\nu} R_{\mu \nu}^{\quad ij},
\end{equation*}
and
\begin{equation*}
    R_{klij} = R_{kl}^{\quad mn} \delta_{mi} \delta_{nj} \\
    R_{ki} = R_{klij} \delta^{lj}.    
\end{equation*}
In the continuous case, we have:
\begin{equation*}
    R_{klij} = R_{ijkl}, \quad R_{kl} = R_{lk}.
\end{equation*}
To test whether the symmetries of the Riemann tensor hold, we examine, in the discrete case, the differences:
\begin{equation}
    R_{1213} - R_{1312}, \quad R_{1223} - R_{2312}, \quad R_{1323} - R_{2313},
\label{differences}
\end{equation}
for the small values of $N$. Figure \ref{R4indices} presents the root-mean-squared error of each of the differences given in equation \ref{differences} versus $N$. This shows that symmetries in the discrete limit do not hold, but the difference is getting smaller as $N$ increases. The curve fitting is $0.5\times N^{-1.97 }$.

\section{Conclusion}

In this paper, we used the definition of the curvature of the discrete space in the recently proposed model of discrete gravity. This is a lattice-like approach. Three integers (positive or negative) are used to enumerate each of the elementary cells. Considering the example of a three-sphere, the curvature was computed numerically. It was shown that as the number of cells increases, the continuous limit is recovered. 

\section*{Appendix I}

Writing the nine equations in detail, we have:
\begin{align}
\mathcal{\ell}T_{\overset{.}{1}\overset{.}{2}}^{1}  &  =0=\widehat{\omega
}_{\overset{.}{1}}^{3}\left(  n\right)  \sin\mathcal{\ell}\omega
_{\overset{.}{1}}\left(  n\right)  e_{\overset{.}{2}}^{2}\left(
n+\overset{.}{1}\right)  +2\widehat{\omega}_{\overset{.}{1}}^{1}\left(
n\right)  \widehat{\omega}_{\overset{.}{1}}^{2}\left(  n\right)  \sin^{2}%
\frac{1}{2}\mathcal{\ell}\omega_{\overset{.}{1}}\left(  n\right)
e_{\overset{.}{2}}^{2}\left(  n+\overset{.}{1}\right) \nonumber\\
&  -\cos\mathcal{\ell}\omega_{\overset{.}{2}}\left(  n\right)
e_{\overset{.}{1}}^{1}\left(  n+\overset{.}{2}\right)  -2\widehat{\omega
}_{\overset{.}{2}}^{1}\left(  n\right)  \widehat{\omega}_{\overset{.}{2}}%
^{1}\left(  n\right)  \sin^{2}\frac{1}{2}\mathcal{\ell}\omega_{\overset{.}{2}%
}\left(  n\right)  e_{\overset{.}{1}}^{1}\left(  n+\overset{.}{2}\right)
+e_{\overset{.}{1}}^{1}\left(  n\right)
\end{align}\label{one}
\begin{align}
\mathcal{\ell}T_{\overset{.}{1}\overset{.}{2}}^{2}  &  =0=\widehat{\omega
}_{\overset{.}{2}}^{3}\left(  n\right)  \sin\mathcal{\ell}\omega
_{\overset{.}{2}}\left(  n\right)  e_{\overset{.}{1}}^{1}\left(
n+\overset{.}{2}\right)  +2\widehat{\omega}_{\overset{.}{1}}^{2}\left(
n\right)  \widehat{\omega}_{\overset{.}{1}}^{2}\left(  n\right)  \sin^{2}%
\frac{1}{2}\mathcal{\ell}\omega_{\overset{.}{1}}\left(  n\right)
e_{\overset{.}{2}}^{2}\left(  n+\overset{.}{1}\right) \nonumber\\
&  +\cos\mathcal{\ell}\omega_{\overset{.}{1}}\left(  n\right)
e_{\overset{.}{2}}^{2}\left(  n+\overset{.}{1}\right)  -2\widehat{\omega
}_{\overset{.}{2}}^{2}\left(  n\right)  \widehat{\omega}_{\overset{.}{2}}%
^{1}\left(  n\right)  \sin^{2}\frac{1}{2}\mathcal{\ell}\omega_{\overset{.}{2}%
}\left(  n\right)  e_{\overset{.}{1}}^{1}\left(  n+\overset{.}{2}\right)
-e_{\overset{.}{2}}^{2}\left(  n\right)
\end{align}%
\begin{align}
\mathcal{\ell}T_{\overset{.}{1}\overset{.}{2}}^{3}  &  =0=-\widehat{\omega
}_{\overset{.}{2}}^{2}\left(  n\right)  \sin\mathcal{\ell}\omega
_{\overset{.}{2}}\left(  n\right)  e_{\overset{.}{1}}^{1}\left(
n+\overset{.}{2}\right)  +2\widehat{\omega}_{\overset{.}{1}}^{2}\left(
n\right)  \widehat{\omega}_{\overset{.}{1}}^{3}\left(  n\right)  \sin^{2}%
\frac{1}{2}\mathcal{\ell}\omega_{\overset{.}{1}}\left(  n\right)
e_{\overset{.}{2}}^{2}\left(  n+\overset{.}{1}\right) \nonumber\\
&  -\widehat{\omega}_{\overset{.}{1}}^{1}\sin\mathcal{\ell}\omega
_{\overset{.}{1}}\left(  n\right)  e_{\overset{.}{2}}^{2}\left(
n+\overset{.}{1}\right)  -2\widehat{\omega}_{\overset{.}{2}}^{1}\left(
n\right)  \widehat{\omega}_{\overset{.}{2}}^{3}\left(  n\right)  \sin^{2}%
\frac{1}{2}\mathcal{\ell}\omega_{\overset{.}{2}}\left(  n\right)
e_{\overset{.}{1}}^{1}\left(  n+\overset{.}{2}\right)
\end{align}%
\begin{align}
\mathcal{\ell}T_{\overset{.}{2}\overset{.}{3}}^{1}  &  =0=-\widehat{\omega
}_{\overset{.}{2}}^{2}\left(  n\right)  \sin\mathcal{\ell}\omega
_{\overset{.}{2}}\left(  n\right)  e_{\overset{.}{3}}^{3}\left(
n+\overset{.}{2}\right)  +2\widehat{\omega}_{\overset{.}{2}}^{1}\left(
n\right)  \widehat{\omega}_{\overset{.}{2}}^{3}\left(  n\right)  \sin^{2}%
\frac{1}{2}\mathcal{\ell}\omega_{\overset{.}{2}}\left(  n\right)
e_{\overset{.}{3}}^{3}\left(  n+\overset{.}{2}\right) \nonumber\\
&  -\widehat{\omega}_{\overset{.}{3}}^{3}\sin\mathcal{\ell}\omega
_{\overset{.}{3}}\left(  n\right)  e_{\overset{.}{2}}^{2}\left(
n+\overset{.}{3}\right)  -2\widehat{\omega}_{\overset{.}{3}}^{1}\left(
n\right)  \widehat{\omega}_{\overset{.}{3}}^{2}\left(  n\right)  \sin^{2}%
\frac{1}{2}\mathcal{\ell}\omega_{\overset{.}{3}}\left(  n\right)
e_{\overset{.}{2}}^{2}\left(  n+\overset{.}{3}\right)
\end{align}%
\begin{align}
\mathcal{\ell}T_{\overset{.}{2}\overset{.}{3}}^{2}  &  =0=\widehat{\omega
}_{\overset{.}{2}}^{1}\left(  n\right)  \sin\mathcal{\ell}\omega
_{\overset{.}{2}}\left(  n\right)  e_{\overset{.}{3}}^{3}\left(
n+\overset{.}{2}\right)  +2\widehat{\omega}_{\overset{.}{2}}^{2}\left(
n\right)  \widehat{\omega}_{\overset{.}{2}}^{3}\left(  n\right)  \sin^{2}%
\frac{1}{2}\mathcal{\ell}\omega_{\overset{.}{2}}\left(  n\right)
e_{\overset{.}{3}}^{3}\left(  n+\overset{.}{2}\right) \nonumber\\
&  -\cos\mathcal{\ell}\omega_{\overset{.}{3}}\left(  n\right)
e_{\overset{.}{2}}^{2}\left(  n+\overset{.}{3}\right)  -2\widehat{\omega
}_{\overset{.}{3}}^{2}\left(  n\right)  \widehat{\omega}_{\overset{.}{3}}%
^{2}\left(  n\right)  \sin^{2}\frac{1}{2}\mathcal{\ell}\omega_{\overset{.}{3}%
}\left(  n\right)  e_{\overset{.}{2}}^{2}\left(  n+\overset{.}{3}\right)
+e_{\overset{.}{2}}^{2}\left(  n\right)
\end{align}%
\begin{align}
\mathcal{\ell}T_{\overset{.}{2}\overset{.}{3}}^{3}  &  =0=\widehat{\omega
}_{\overset{.}{3}}^{1}\left(  n\right)  \sin\mathcal{\ell}\omega
_{\overset{.}{3}}\left(  n\right)  e_{\overset{.}{2}}^{2}\left(
n+\overset{.}{3}\right)  +2\widehat{\omega}_{\overset{.}{2}}^{3}\left(
n\right)  \widehat{\omega}_{\overset{.}{2}}^{3}\left(  n\right)  \sin^{2}%
\frac{1}{2}\mathcal{\ell}\omega_{\overset{.}{2}}\left(  n\right)
e_{\overset{.}{3}}^{3}\left(  n+\overset{.}{2}\right) \nonumber\\
&  +\cos\mathcal{\ell}\omega_{\overset{.}{2}}\left(  n\right)
e_{\overset{.}{3}}^{3}\left(  n+\overset{.}{2}\right)  -2\widehat{\omega
}_{\overset{.}{3}}^{3}\left(  n\right)  \widehat{\omega}_{\overset{.}{3}}%
^{2}\left(  n\right)  \sin^{2}\frac{1}{2}\mathcal{\ell}\omega_{\overset{.}{3}%
}\left(  n\right)  e_{\overset{.}{2}}^{2}\left(  n+\overset{.}{3}\right)
-e_{\overset{.}{3}}^{3}\left(  n\right)
\end{align}%
\begin{align}
\mathcal{\ell}T_{\overset{.}{1}\overset{.}{3}}^{1}  &  =0=-\widehat{\omega
}_{\overset{.}{1}}^{2}\left(  n\right)  \sin\mathcal{\ell}\omega
_{\overset{.}{1}}\left(  n\right)  e_{\overset{.}{3}}^{3}\left(
n+\overset{.}{1}\right)  +2\widehat{\omega}_{\overset{.}{1}}^{1}\left(
n\right)  \widehat{\omega}_{\overset{.}{1}}^{3}\left(  n\right)  \sin^{2}%
\frac{1}{2}\mathcal{\ell}\omega_{\overset{.}{1}}\left(  n\right)
e_{\overset{.}{3}}^{3}\left(  n+\overset{.}{1}\right) \nonumber\\
&  -\cos\mathcal{\ell}\omega_{\overset{.}{3}}\left(  n\right)
e_{\overset{.}{1}}^{1}\left(  n+\overset{.}{3}\right)  -2\widehat{\omega
}_{\overset{.}{3}}^{1}\left(  n\right)  \widehat{\omega}_{\overset{.}{3}}%
^{1}\left(  n\right)  \sin^{2}\frac{1}{2}\mathcal{\ell}\omega_{\overset{.}{3}%
}\left(  n\right)  e_{\overset{.}{1}}^{1}\left(  n+\overset{.}{3}\right)
+e_{\overset{.}{1}}^{1}\left(  n\right)
\end{align}%
\begin{align}
\mathcal{\ell}T_{\overset{.}{1}\overset{.}{3}}^{2}  &  =0=\widehat{\omega
}_{\overset{.}{1}}^{1}\left(  n\right)  \sin\mathcal{\ell}\omega
_{\overset{.}{1}}\left(  n\right)  e_{\overset{.}{3}}^{3}\left(
n+\overset{.}{1}\right)  +2\widehat{\omega}_{\overset{.}{1}}^{2}\left(
n\right)  \widehat{\omega}_{\overset{.}{1}}^{3}\left(  n\right)  \sin^{2}%
\frac{1}{2}\mathcal{\ell}\omega_{\overset{.}{1}}\left(  n\right)
e_{\overset{.}{3}}^{3}\left(  n+\overset{.}{1}\right) \nonumber\\
&  +\widehat{\omega}_{\overset{.}{3}}^{3}\sin\mathcal{\ell}\omega
_{\overset{.}{3}}\left(  n\right)  e_{\overset{.}{1}}^{1}\left(
n+\overset{.}{3}\right)  -2\widehat{\omega}_{\overset{.}{3}}^{1}\left(
n\right)  \widehat{\omega}_{\overset{.}{3}}^{2}\left(  n\right)  \sin^{2}%
\frac{1}{2}\mathcal{\ell}\omega_{\overset{.}{3}}\left(  n\right)
e_{\overset{.}{1}}^{1}\left(  n+\overset{.}{3}\right)
\end{align}%
\begin{align}
\mathcal{\ell}T_{\overset{.}{1}\overset{.}{3}}^{3}  &  =0=-\widehat{\omega
}_{\overset{.}{3}}^{2}\left(  n\right)  \sin\mathcal{\ell}\omega
_{\overset{.}{3}}\left(  n\right)  e_{\overset{.}{1}}^{1}\left(
n+\overset{.}{3}\right)  +2\widehat{\omega}_{\overset{.}{1}}^{3}\left(
n\right)  \widehat{\omega}_{\overset{.}{1}}^{3}\left(  n\right)  \sin^{2}%
\frac{1}{2}\mathcal{\ell}\omega_{\overset{.}{1}}\left(  n\right)
e_{\overset{.}{3}}^{3}\left(  n+\overset{.}{1}\right) \nonumber\\
&  +\cos\mathcal{\ell}\omega_{\overset{.}{1}}\left(  n\right)
e_{\overset{.}{3}}^{3}\left(  n+\overset{.}{1}\right)  -2\widehat{\omega
}_{\overset{.}{3}}^{3}\left(  n\right)  \widehat{\omega}_{\overset{.}{3}}%
^{1}\left(  n\right)  \sin^{2}\frac{1}{2}\mathcal{\ell}\omega_{\overset{.}{3}%
}\left(  n\right)  e_{\overset{.}{1}}^{1}\left(  n+\overset{.}{3}\right)
-e_{\overset{.}{3}}^{3}\left(  n\right)
\end{align} \label{nine}
In this notation we have $e_{\overset{.}{1}}^{1}\left(  n+\overset{.}{2}%
\right)  \equiv e_{\overset{.}{1}}^{1}\left(  n_{1},n_{2}+1,n_{3}\right)  $
thus shifting only the second coordinate.

\section*{Appendix II}

\begin{align}
R_{\overset{.}{1}\overset{.}{2}}^{\quad1}\left(  n\right)   &  =\frac
{2}{\mathcal{\ell}^{2}}\left(  A_{\overset{.}{1}\overset{.}{2}}%
B_{\overset{.}{1}\overset{.}{2}}^{1}-A_{\overset{.}{2}\overset{.}{1}%
}B_{\overset{.}{2}\overset{.}{1}}^{1}+B_{\overset{.}{1}\overset{.}{2}}%
^{2}\left(  n\right)  B_{\overset{.}{2}\overset{.}{1}}^{3}\left(  n\right)
-B_{\overset{.}{1}\overset{.}{2}}^{3}\left(  n\right)  B_{\overset{.}{2}%
\overset{.}{1}}^{2}\left(  n\right)  \right) \\
R_{\overset{.}{1}\overset{.}{2}}^{\quad2}\left(  n\right)   &  =\frac
{2}{\mathcal{\ell}^{2}}\left(  A_{\overset{.}{1}\overset{.}{2}}%
B_{\overset{.}{1}\overset{.}{2}}^{2}-A_{\overset{.}{2}\overset{.}{1}%
}B_{\overset{.}{2}\overset{.}{1}}^{2}+B_{\overset{.}{1}\overset{.}{2}}%
^{3}\left(  n\right)  B_{\overset{.}{2}\overset{.}{1}}^{1}\left(  n\right)
-B_{\overset{.}{1}\overset{.}{2}}^{1}\left(  n\right)  B_{\overset{.}{2}%
\overset{.}{1}}^{3}\left(  n\right)  \right) \\
R_{\overset{.}{1}\overset{.}{2}}^{\quad3}\left(  n\right)   &  =\frac
{2}{\mathcal{\ell}^{2}}\left(  A_{\overset{.}{1}\overset{.}{2}}%
B_{\overset{.}{1}\overset{.}{2}}^{3}-A_{\overset{.}{2}\overset{.}{1}%
}B_{\overset{.}{2}\overset{.}{1}}^{3}+B_{\overset{.}{1}\overset{.}{2}}%
^{1}\left(  n\right)  B_{\overset{.}{2}\overset{.}{1}}^{2}\left(  n\right)
-B_{\overset{.}{1}\overset{.}{2}}^{2}\left(  n\right)  B_{\overset{.}{2}%
\overset{.}{1}}^{1}\left(  n\right)  \right)
\end{align}%
\begin{align}
R_{\overset{.}{2}\overset{.}{3}}^{\quad1}\left(  n\right)   &  =\frac
{2}{\mathcal{\ell}^{2}}\left(  A_{\overset{.}{2}\overset{.}{3}}%
B_{\overset{.}{2}\overset{.}{3}}^{1}-A_{\overset{.}{3}\overset{.}{2}%
}B_{\overset{.}{3}\overset{.}{2}}^{1}+B_{\overset{.}{2}\overset{.}{3}}%
^{2}\left(  n\right)  B_{\overset{.}{3}\overset{.}{2}}^{3}\left(  n\right)
-B_{\overset{.}{2}\overset{.}{3}}^{3}\left(  n\right)  B_{\overset{.}{3}%
\overset{.}{2}}^{2}\left(  n\right)  \right) \\
R_{\overset{.}{2}\overset{.}{3}}^{\quad2}\left(  n\right)   &  =\frac
{2}{\mathcal{\ell}^{2}}\left(  A_{\overset{.}{2}\overset{.}{3}}%
B_{\overset{.}{2}\overset{.}{3}}^{2}-A_{\overset{.}{3}\overset{.}{2}%
}B_{\overset{.}{3}\overset{.}{2}}^{2}+B_{\overset{.}{2}\overset{.}{3}}%
^{3}\left(  n\right)  B_{\overset{.}{3}\overset{.}{2}}^{1}\left(  n\right)
-B_{\overset{.}{2}\overset{.}{3}}^{1}\left(  n\right)  B_{\overset{.}{3}%
\overset{.}{2}}^{3}\left(  n\right)  \right) \\
R_{\overset{.}{2}\overset{.}{3}}^{\quad3}\left(  n\right)   &  =\frac
{2}{\mathcal{\ell}^{2}}\left(  A_{\overset{.}{2}\overset{.}{3}}%
B_{\overset{.}{2}\overset{.}{3}}^{3}-A_{\overset{.}{3}\overset{.}{2}%
}B_{\overset{.}{3}\overset{.}{2}}^{3}+B_{\overset{.}{2}\overset{.}{3}}%
^{1}\left(  n\right)  B_{\overset{.}{3}\overset{.}{2}}^{2}\left(  n\right)
-B_{\overset{.}{2}\overset{.}{3}}^{2}\left(  n\right)  B_{\overset{.}{3}%
\overset{.}{2}}^{1}\left(  n\right)  \right)
\end{align}%
\begin{align}
R_{\overset{.}{1}\overset{.}{3}}^{\quad1}\left(  n\right)   &  =\frac
{2}{\mathcal{\ell}^{2}}\left(  A_{\overset{.}{1}\overset{.}{3}}%
B_{\overset{.}{1}\overset{.}{3}}^{1}-A_{\overset{.}{3}\overset{.}{1}%
}B_{\overset{.}{3}\overset{.}{1}}^{1}+B_{\overset{.}{1}\overset{.}{3}}%
^{2}\left(  n\right)  B_{\overset{.}{3}\overset{.}{1}}^{3}\left(  n\right)
-B_{\overset{.}{1}\overset{.}{3}}^{3}\left(  n\right)  B_{\overset{.}{3}%
\overset{.}{1}}^{2}\left(  n\right)  \right) \\
R_{\overset{.}{1}\overset{.}{3}}^{\quad2}\left(  n\right)   &  =\frac
{2}{\mathcal{\ell}^{2}}\left(  A_{\overset{.}{1}\overset{.}{3}}%
B_{\overset{.}{1}\overset{.}{3}}^{2}-A_{\overset{.}{3}\overset{.}{1}%
}B_{\overset{.}{3}\overset{.}{1}}^{2}+B_{\overset{.}{1}\overset{.}{3}}%
^{3}\left(  n\right)  B_{\overset{.}{3}\overset{.}{1}}^{1}\left(  n\right)
-B_{\overset{.}{1}\overset{.}{3}}^{1}\left(  n\right)  B_{\overset{.}{3}%
\overset{.}{1}}^{3}\left(  n\right)  \right) \\
R_{\overset{.}{1}\overset{.}{3}}^{\quad3}\left(  n\right)   &  =\frac
{2}{\mathcal{\ell}^{2}}\left(  A_{\overset{.}{1}\overset{.}{3}}%
B_{\overset{.}{1}\overset{.}{3}}^{3}-A_{\overset{.}{3}\overset{.}{1}%
}B_{\overset{.}{3}\overset{.}{1}}^{3}+B_{\overset{.}{1}\overset{.}{3}}%
^{1}\left(  n\right)  B_{\overset{.}{3}\overset{.}{1}}^{2}\left(  n\right)
-B_{\overset{.}{1}\overset{.}{3}}^{2}\left(  n\right)  B_{\overset{.}{3}%
\overset{.}{1}}^{1}\left(  n\right)  \right)
\end{align}
We will also give an example of how the components of $A_{\mu\nu}\left(
n\right)  $ and $B_{\mu\nu}^{i}\left(  n\right)  $ look like%
\begin{align}
A_{\overset{.}{1}\overset{.}{2}}  &  =\cos\frac{1}{2}\mathcal{\ell}%
\omega_{\overset{.}{1}}\left(  n+\overset{.}{2}\right)  \cos\frac{1}%
{2}\mathcal{\ell}\omega_{\overset{.}{2}}\left(  n\right)  -%
{\displaystyle\sum\limits_{j=1}^{3}}
\widehat{\omega}_{\overset{.}{1}}^{j}\left(  n+\overset{.}{2}\right)
\widehat{\omega}_{\overset{.}{2}}^{j}\left(  n\right)  \sin\frac{1}%
{2}\mathcal{\ell}\omega_{\overset{.}{1}}\left(  n+\overset{.}{2}\right)
\sin\frac{1}{2}\mathcal{\ell}\omega_{\overset{.}{2}}\left(  n\right) \\
A_{\overset{.}{2}\overset{.}{1}}  &  =\cos\frac{1}{2}\mathcal{\ell}%
\omega_{\overset{.}{2}}\left(  n+\overset{.}{1}\right)  \cos\frac{1}%
{2}\mathcal{\ell}\omega_{\overset{.}{1}}\left(  n\right)  -%
{\displaystyle\sum\limits_{j=1}^{3}}
\widehat{\omega}_{\overset{.}{2}}^{j}\left(  n+\overset{.}{2}\right)
\widehat{\omega}_{\overset{.}{1}}^{j}\left(  n\right)  \sin\frac{1}%
{2}\mathcal{\ell}\omega_{\overset{.}{2}}\left(  n+\overset{.}{2}\right)
\sin\frac{1}{2}\mathcal{\ell}\omega_{\overset{.}{1}}\left(  n\right)
\end{align}%
\begin{align}
B_{\overset{.}{1}\overset{.}{2}}^{i}\left(  n\right)   &  =\left(
\widehat{\omega}_{\overset{.}{1}}^{i}\left(  n\right)  \sin\frac{1}%
{2}\mathcal{\ell}\omega_{\overset{.}{1}}\left(  n\right)  \cos\frac{1}%
{2}\mathcal{\ell}\omega_{\overset{.}{2}}\left(  n+\overset{.}{1}\right)
+\widehat{\omega}_{\overset{.}{2}}^{i}\left(  n+\overset{.}{1}\right)
\sin\frac{1}{2}\mathcal{\ell}\omega_{\overset{.}{2}}\left(  n+\overset{.}{1}%
\right)  \cos\frac{1}{2}\mathcal{\ell}\omega_{\overset{.}{1}}\left(  n\right)
\right. \nonumber\\
&  \left.  -\epsilon^{ijk}\widehat{\omega}_{\overset{.}{1}}^{j}\left(
n\right)  \sin\frac{1}{2}\mathcal{\ell}\omega_{\overset{.}{1}}\left(
n\right)  \widehat{\omega}_{\overset{.}{2}}^{k}\left(  n+\overset{.}{1}%
\right)  \sin\frac{1}{2}\mathcal{\ell}\omega_{\overset{.}{2}}\left(
n+\overset{.}{1}\right)  \right)
\end{align}

\section*{Acknowledgments}
A.H.C would like to thank Slava Mukhanov for suggesting the use of isotropic coordinates. The work of A. H. C is supported in part by the National Science Foundation Grant No. Phys-2207663. 

\section*{Supporting Material}
The script used in the paper is available at the following 

\href{https://www.dropbox.com/s/mgiqmh2ukfapny1/3DcaseFinalCode.R?dl=0}{link.}

\bibliography{references}

\end{document}